\documentclass[superscriptaddress,secnumarabic,nobibnotes,aps,prd,showkeys,showpacs,12pt, nofootinbib]{revtex4-1}

\usepackage[nointlimits]{amsmath}
\usepackage{hyperref}
\usepackage{amsfonts}
\usepackage{amssymb}
\usepackage{amsmath}
\usepackage{amsthm}
\usepackage{latexsym}
\usepackage{graphicx}
\usepackage{color}
\usepackage{layout}
\usepackage[english]{babel}
\usepackage[utf8]{inputenc}

\allowdisplaybreaks

\begin{document}

\title{Nonminimally coupled Weyl gravity}

\author{Cláudio Gomes}
\email{claudio.gomes@fc.up.pt}
\affiliation{Departamento de Física e Astronomia, Faculdade de Ciências da Universidade do Porto \\ and Centro de Física do Porto \\ Rua do Campo Alegre 687, 4169-007, Porto, Portugal}

\author{Orfeu Bertolami}
\email{orfeu.bertolami@fc.up.pt}
\affiliation{Departamento de Física e Astronomia, Faculdade de Ciências da Universidade do Porto \\ and Centro de Física do Porto \\ Rua do Campo Alegre 687, 4169-007, Porto, Portugal}


\date{\today}

\begin{abstract}
Weyl gravity in the presence of a non-minimal matter-curvature coupling is presented. Some properties arising from the non-metricity give rise to a second order theory whose vacuum is compatible with a cosmological constant, under conditions on the vector field.
\end{abstract}


\maketitle

\section{Introduction}

It is well known that the most general affine connection can be decomposed as a sum of three different parts: the Christoffel symbols (built from the metric), the antisymmetric distortion tensor and the disformation tensor associated to non-metricity \cite{connection1}. In General Relativity (GR) only the first term is nonvanishing and thus the connection is the Levi-Civita one. Nevertheless, the other terms could be nonzero leading to other theories of gravity. Some extensions of GR assume either just torsion \cite{teleparallel} or only non-metricity \cite{symmetricteleparallel, nonmetricityGR,nonmetricityGR2}. In these extensions many new observational features emerge. There is a large literature on theories with torsion (See Refs. \cite{torsionreview1,torsionreview2} for reviews). In what concerns non-metricity, it was, for instance, constrained by the Bhabha scattering in the metric-affine formulation of Ricci-based gravity models in vacuum  \cite{nonmetricitytests}, and a lower bound on the scale of non-metricity was found to be greater than $1~ TeV$.

Although GR can account accurately for observational data at solar system, there are several reasons to seek for alternative gravitational models. These include the difficulty in achieving a fully consistent UV-completion of the theory and the existence of large scale deviations, fixed by the inclusion of two unknown dark components, dark matter and dark energy, which have not been directly detected yet. Therefore, many extensions have been proposed: f(R) theories \cite{felice,soutiriou} and non-minimal matter-curvature coupling \cite{Rcoupling,nmc} among others.

In particular, the non-minimal matter-curvature coupling theories (NMC) were shown to have interesting observational implications. For instance, they mimic dark matter at galaxies and clusters \cite{dmgalaxies,dmclusters}, and also dark energy at large scales \cite{demim,curraccel}. They modify the Layer-Irvine equation and the virial theorem at cluster scales \cite{li}, and have bearings on black hole solutions \cite{bh}, on scalar field inflation and preheating \cite{inflation, reheating} and on gravitational waves \cite{nmcgw}. Furthermore, these theories were shown to be stable with respect to cosmological perturbations \cite{frazao} and to the Dolgov-Kawasaki criterion \cite{Bertolami:2009cd}. These theories also admit a Palatini formulation \cite{nmcpalatini}.

There have been also some work on theories that extended the effects of torsion, such as $f(T)$ \cite{fT1,fT2}, or of non-metricity \cite{nonmetricityfR}, which differ significantly from $f(R)$ theories. The differences if $f(R)=R$ are not so sharp \cite{equivalence}. Non-minimal coupled versions with curvature and torsion \cite{fRT} and with non-metricity \cite{nmcQ} have been considered. One class of non-metric theories is the Weyl gravity \cite{weyl}, which in its original form was proposed to unify gravity and electromagnetism. Later on, there was a reformulation due to Dirac leading to a simpler action, but that included a new scalar field to describe spacetime in addition to the metric \cite{dirac}. Recently, there has been a revival of Weyl gravity-like theories as solutions for the dark matter and dark energy problems or inflation \cite{weylgravity}.

Actually, the interest of vector fields in cosmology is not new \cite{vector, vector2, vector3, vector5}. A generic vector may destroy the homogeneity and isotropy of the Universe, unless if it is embedded in a $SO(3)$ symmetry \cite{vector2,vector3,vector5}.

The main goal of the present work is to discuss Weyl gravity with a non-minimal coupling between matter and curvature. The properties of this model are discussed and some implications for cosmology are examined. This paper is organised as follows. In Section \ref{sec:nmc} we review the non-minimal matter-curvature model and some of its properties. In Section \ref{sec:weyl}, we look into Weyl gravity and discuss its main geometric features. In Section \ref{sec:weylnmc}, we explore the main properties of Weyl gravity generalised so to allow for the non-minimal coupling between matter and curvature. We further discuss in Section \ref{sec:sequestering} the use of the generalised Bianchi identities to consider a cosmological constant arising from the theory. We study the space form solution and require Weyl vector field to be compatible with a homogeneous and isotropic Universe and a time varying cosmological term in Section \ref{sec:spaceform}. We draw our conclusions in Section \ref{sec:conclusions}.

\section{Nonminimal matter-curvature coupling}\label{sec:nmc}

The action functional for theories with non-minimal matter-curvature coupling can be expressed as:
\begin{equation}
S=\int \left(\kappa f_1(R)+f_2(R)\mathcal{L}\right) \sqrt{-g} d^4x ~,
\end{equation}
where both $f_1(R), f_2(R)$ are generic functions of the scalar curvature, $\kappa=1/16\pi G$ with $G$ being the Newton's constant, $\mathcal{L}$ is the matter Lagrangian density and $g$ is the determinant of the metric, $g_{\mu\nu}$.

Varying the action with respect to the metric yields:
\begin{equation}
\Theta R_{\mu\nu}-\frac{1}{2}g_{\mu\nu}f_1=\frac{f_2}{2\kappa}T_{\mu\nu}+\square_{\mu\nu}\Theta~,
\end{equation}
where we have defined $\Theta :=F_1+(F_2/\kappa)\mathcal{L}$, with $F_i:= df_i/dR$, $T_{\mu\nu}$ is the usual energy-momentum tensor built from the matter Lagrangian, and we have also defined the second order operator $\square_{\mu\nu}:=\nabla_{\mu}\nabla_{\nu}-g_{\mu\nu}\square$.

One of the most striking features of this model is the covariant non-conservation of the energy-momentum tensor:
\begin{equation}
\nabla_{\mu}T^{\mu\nu}=\frac{F_2}{f_2}\left(g^{\mu\nu}\mathcal{L}-T^{\mu\nu}\right)\nabla_{\mu}R~.
\end{equation}

This feature has several implications. One of the non trivial ones is the lifting of the degeneracy in the choice of the matter Lagrangian density that leads to a perfect fluid energy-momentum tensor (see Ref. \cite{perfectfluid} for a thorough discussion).

Until now, we have considered that the connection is metric compatible and torsionless. However, we can add non-metricity to the connection, and a particular interesting case of such extensions is the Weyl Gravity.

\section{Weyl Gravity}\label{sec:weyl}

Weyl gravity is based on the existence of a vector field which introduces non-metric properties to the connection. This can be expressed through a generalised covariant derivative\footnote{Notice that the covariant derivative of the inverse metric is given by:
$D_{\lambda}g^{\mu\nu}=-A_{\lambda}g^{\mu\nu}$.}
\begin{equation}
D_{\lambda}g_{\mu\nu}=A_{\lambda}g_{\mu\nu}~,
\end{equation}
where,
\begin{equation}
D_{\lambda}g_{\mu\nu}=\nabla_{\lambda}g_{\mu\nu}-\bar{\bar{\Gamma}}^{\rho}_{\mu\lambda}g_{\rho\nu}-\bar{\bar{\Gamma}}^{\rho}_{\nu\lambda}g_{\rho\mu}~,
\end{equation}
 $\nabla_{\lambda}$ is the usual covariant derivative with Levi-Civita connection and $\bar{\bar{\Gamma}}^{\rho}_{\mu\nu}=-\frac{1}{2}\delta^{\rho}_{\mu}A_{\nu}-\frac{1}{2}\delta^{\rho}_{\nu}A_{\mu}+\frac{1}{2}g_{\mu\nu}A^{\rho}$ is the Weyl connection which reflects the non-metricity.

For the Weyl connection, the contraction of the Riemann tensor, the Ricci tensor, $\bar{R}_{\mu\nu}:= \bar{R}^{\lambda}_{\mu\lambda\nu}$, is given by:
\begin{equation}
\bar{R}_{\mu\nu}=R_{\mu\nu}+\frac{1}{2}A_{\mu}A_{\nu}+\frac{1}{2}g_{\mu\nu}\left(\nabla_{\lambda}-A_{\lambda}\right)A^{\lambda}+F_{\mu\nu}+\frac{1}{2}\left(\nabla_{\mu}A_{\nu}+\nabla_{\nu}A_{\mu}\right):=R_{\mu\nu}+\bar{\bar{R}}_{\mu\nu}~,
\end{equation}
where $F_{\mu\nu}:=\partial_{\mu}A_{\nu}-\partial_{\nu}A_{\mu}=\nabla_{\mu}A_{\nu}-\nabla_{\nu}A_{\mu}$ is the strength tensor of the Weyl field. The contraction of the Ricci tensor gives the scalar curvature:
\begin{equation}
\bar{R}=R+3\nabla_{\lambda}A^{\lambda}-\frac{3}{2}A_{\lambda}A^{\lambda}:=R+\bar{\bar{R}}~.
\end{equation}

Therefore, the spacetime is not specified just by the metric, but also by the vector field. Therefore, the constant Riemann curvature solutions also depend on the values of the vector field and only coincide with the usual spaces (de Sitter, anti-de Sitter or Minkowski spaces) when the Weyl vector vanishes everywhere. 

The contracted Bianchi identities with the Weyl covariant derivative read:
\begin{equation}
\mathcal{D}_{\mu}\bar{G}^{\mu\nu}=-\frac{1}{2}\mathcal{D}_{\mu}F^{\mu\nu}~,
\end{equation}
which give:
\begin{eqnarray}
&&\mathcal{D}_{\mu}\bar{G}^{(\mu\nu)}=0 ~,\\
&&\mathcal{D}_{\mu}F^{\mu\nu}=0~,
\end{eqnarray}
where $\bar{G}_{\mu\nu}=\bar{R}_{\mu\nu}-\frac{1}{2}g_{\mu\nu}\bar{R}$ is the Einstein-like tensor for the $\bar{R}^{(\mu\nu)}$ curvature and we have defined $\mathcal{D}_{\mu}:=\left(D_{\mu}+2A_{\mu}\right)$.

We can generalise Weyl gravity by considering the non-minimal matter-curvature coupling of Ref. \cite{nmc}.

\section{Weyl Gravity with a non-minimal matter-curvature coupling}\label{sec:weylnmc}

We now consider the non-minimal coupling model with a Weyl connection. The action functional should read:
\begin{equation}
S=\int \left(\kappa f_1(\bar{R})+f_2(\bar{R})\mathcal{L}\right) \sqrt{-g} d^4x ~.
\end{equation}

Varying the action with respect to the vector field yields, up to boundary terms, a constraint equation:
\begin{equation}
\nabla_{\lambda}\bar{\Theta}=-A_{\lambda}\bar{\Theta} ~,
\end{equation}
where $\bar{\Theta}=F_1(\bar{R})+(F_2(\bar{R})/\kappa)\mathcal{L}$, and $F_i:=df_i/d\bar{R}$. If we vary the action with respect to the metric and taking into account the previous equation, we find:
\begin{equation}
\left[R_{\mu\nu}+\bar{\bar{R}}_{(\mu\nu)}\right]\bar{\Theta}-\frac{1}{2}g_{\mu\nu}f_1=\frac{f_2}{2\kappa}T_{\mu\nu}~,
\label{eqn:metricfield}
\end{equation}
where $\bar{\bar{R}}_{(\mu\nu)}=\frac{1}{2}A_{\mu}A_{\nu}+\frac{1}{2}g_{\mu\nu}\left(\nabla_{\lambda}-A_{\lambda}\right)A^{\lambda}+\nabla_{(\mu}A_{\nu)}$.

The constraint equation for the vector field reduces a fourth order theory, as the usual NMC, into a second order version.

The trace of the metric field equations reads:
\begin{equation}
\bar{R}\bar{\Theta}-2f_1=\frac{f_2}{2\kappa}T~.
\end{equation}
 
Taking the trace of the metric field equations and substituting into Eq. (\ref{eqn:metricfield}), we have the trace-free equations:
\begin{equation}
\bar{\Theta}\left[R_{\mu\nu}-\frac{1}{4}g_{\mu\nu}R\right]+\bar{\Theta}\left(\bar{\bar{R}}_{(\mu\nu)}-\frac{1}{4}g_{\mu\nu}\bar{\bar{R}}\right)=\frac{f_2}{2\kappa}\left[{T}_{\mu\nu}-\frac{1}{4}g_{\mu\nu}{T}\right] ~,
\end{equation}

Taking the divergence of the field equations leads to the covariant non-conservation law for the energy-momentum tensor:
\begin{equation}
\nabla_{\mu}T^{\mu\nu}=\frac{2}{f_2}\left[\frac{F_2}{2}\left(g^{\mu\nu}\mathcal{L}-T^{\mu\nu}\right)\nabla_{\mu}R+\nabla_{\mu}\left(\bar{\Theta B^{\mu\nu}}\right)-\frac{1}{2}\left(F_1g^{\mu\nu}+F_2T^{\mu\nu}\right)\nabla_{\mu}\bar{\bar{R}}\right]~,
\end{equation}
where we have defined the tensor: $B^{\mu\nu}:=\frac{3}{2}A^{\mu}A^{\nu}+\frac{3}{2}g^{\mu\nu}\left(\nabla_{\lambda}-A_{\lambda}\right)A^{\lambda}$.

On its hand, the Weyl divergence of the energy-momentum tensor is given by:
\begin{equation}
\mathcal{D}_{\mu}T^{\mu\nu}=\frac{2\bar{\Theta}}{f_2}\left[\mathcal{D}_{\mu}\bar{R}^{(\mu\nu)}-\frac{1}{2}\mathcal{D}_{\mu}\left(g^{\mu\nu}\frac{f_1}{\bar{\Theta}}\right)-\mathcal{D}_{\mu}\left(\frac{f_2}{2\bar{\Theta}}\right)T^{\mu\nu}\right]~,
\end{equation}
where we have not further computed the resulting expression since this form will be useful in the next section when considering space form for the Riemann curvature.

\subsection{The sequestering of the cosmological constant}\label{sec:sequestering}

In the spirit of unimodular gravity \cite{unimodular1,unimodular2,unimodular3,unimodular4,unimodular5} and of the relaxed regime for the non-minimal coupling model \cite{sequestering} (based on the proposal of Refs. \cite{sequestering1,sequestering2}) we aim to obtain an integration constant from the Bianchi identities for the non-minimally coupled Weyl gravity. Let us make $\kappa=1$ and bearing in mind the relation $\left(\square\nabla^{\nu}-\nabla^{\nu}\square\right)H=R^{\mu\nu}\nabla_{\mu}H$, for a scalar function $H$, we can compute:
\begin{eqnarray}
\frac{1}{4}\nabla^{\nu}R&&=\nabla_{\mu}\left(R^{\mu\nu}-\frac{1}{4}g^{\mu\nu}R\right)
 \nonumber\\
&&=\frac{1}{4}\nabla^{\nu}R+\nabla_{\mu}\left(\frac{1}{\bar{\Theta}}\right)\left[-(R^{\mu\nu}+B^{\mu\nu})\bar{\Theta}-\frac{1}{2}g^{\mu\nu}f_1+\frac{f_2}{2}T^{\mu\nu}+\square^{\mu\nu}\bar{\Theta}\right]+\nonumber\\
&&+\frac{1}{4}\nabla^{\nu}\left[R+B-2f_1-\frac{f_2}{2\bar{\Theta}}+\frac{3}{\bar{\Theta}}\square\bar{\Theta}\right]\\
&&\iff \nabla^{\nu}\left[R+B-2f_1-\frac{f_2}{2\bar{\Theta}}+\frac{3}{\bar{\Theta}}\square\bar{\Theta}\right]=0~,
\end{eqnarray}
as obtained from the covariant derivative of the trace of the field equations. Therefore, as in Ref. \cite{sequestering}, the Bianchi identities do not provide an integration constant which could be identified as the cosmological constant.

Analogously, we can attempt to repeat the procedure for the generalised Bianchi identities:
\begin{eqnarray}
&&\frac{1}{4}\mathcal{D}_{\mu}\left(g^{\mu\nu}\bar{R}\right)=\mathcal{D}_{\mu}\left[\bar{R}^{(\mu\nu)}-\frac{1}{4}g^{\mu\nu}\bar{R}\right]=\mathcal{D}_{\mu}\left[\frac{f_2}{2\bar{\Theta}}\left(T^{\mu\nu}-\frac{1}{4}g^{\mu\nu}T\right)\right]=\nonumber\\
&&=\mathcal{D}_{\mu}\left(\frac{f_2}{2\bar{\Theta}}\right)T^{\mu\nu}+\left[\mathcal{D}_{\mu}\bar{R}^{(\mu\nu)}-\frac{1}{2}\mathcal{D}_{\mu}\left(g^{\mu\nu}\frac{f_1}{\bar{\Theta}}\right)-\mathcal{D}_{\mu}\left(\frac{f_2}{2\bar{\Theta}}\right)T^{\mu\nu}\right]-\frac{1}{4}\mathcal{D}_{\mu}\left(g^{\mu\nu}\frac{f_2}{2\bar{\Theta}}T\right)\\
&&\iff \mathcal{D}_{\mu}\left(\bar{R}^{(\mu\nu)}-\frac{1}{4}g^{\mu\nu}\bar{R}\right)+\frac{1}{4}\mathcal{D}_{\mu}\left[g^{\mu\nu}\left(-\frac{f_2}{2\bar{\Theta}}T-2\frac{f_1}{\bar{\Theta}}\right)\right]=0\\
&&\iff \frac{1}{4}\mathcal{D}_{\mu}\left[g^{\mu\nu}\left(\bar{R}-\frac{f_2}{2\bar{\Theta}}T-2\frac{f_1}{\bar{\Theta}}\right)\right]=0~,
\end{eqnarray}
where the expression to be differentiated is simply the trace of the metric equations, which, of course, do not imply that there is an integration constant. 

However, we may wonder whether the non-minimal coupling can admit constant sectional curvature solutions. We shall explore this possibility in the next section.

\subsection{The space form behaviour}\label{sec:spaceform}

We now aim to assess whether the vacuum of the theory admits a constant generalised curvature solution. For that, let us recall that a (pseudo-Riemannian) manifold is said to be a space form if and only if \cite{spaceform}
\begin{equation}
\bar{R}_{abcd}=K(g_{ac}g_{db}-g_{ad}g_{cb})\implies\bar{R}_{bd}=3Kg_{db}\implies\bar{R}=12K~,
\end{equation}
where $K$ is some real constant, which in GR is directly related to the cosmological constant up to a numerical factor, and characterises a well defined vacuum state for the gravity theory. This can, however, be generalised to the case of homogeneous and isotropic evolving spaces where $K=K(t)$ \footnote{One homogeneous and isotropic, but time varying scalar curvature leads to a time varying cosmological term that might reproduce the dark energy behaviour of the Universe. In fact, the time variation of a cosmological term was proposed in the context of the Brans-Dick theory in Ref. \cite{varyingcosmologicalconstant}.}.

Considering the contribution from the vacuum to the Lagrangian density, $\mathcal{L}=-2\kappa\Lambda_0$, the metric field equations and their trace give:
\begin{eqnarray}
&&\left[R_{\mu\nu}+\bar{\bar{R}}_{(\mu\nu)}\right]\bar{\Theta}=\frac{1}{2}g_{\mu\nu}f_1-g_{\mu\nu}f_2\Lambda_0~,\label{eqn:fieldandtrace1}\\
&&\left[R+\bar{\bar{R}}\right]\bar{\Theta}=2f_1-4f_2\Lambda_0~.\label{eqn:fieldandtrace2}
\end{eqnarray}

Rearranging the trace of the metric field equations, Eq. (\ref{eqn:fieldandtrace2}):
\begin{equation}
\frac{2f_1-\bar{R}F_1}{2f_2-\bar{R}F_2}=2\Lambda_0~,
\end{equation}
which may allow for a way to sequester the cosmological constant. As an example, if we consider that  $f_1=\bar{R}$, then:
\begin{equation}
f_2(\bar{R})=\frac{\bar{R}}{2\Lambda_0}+\mathcal{C}_2\bar{R}^2 \iff f_2(\bar{R}=12K(t)) = \frac{6K(t)}{\Lambda_0}+\mathcal{C}'_2K(t)^2~,
\end{equation}
where $\mathcal{C}_2$ is an integration constant and $\mathcal{C}'_2=144\mathcal{C}_2$.

If instead, we have a feeble non-minimal coupling $f_2(\bar{R})=1$, but the pure gravity does not reduce to GR behaviour, then:
\begin{equation}
f_1(\bar{R})=2\Lambda_0+\mathcal{C}_1\bar{R}^2\iff f_1(\bar{R}=12K(t))=2\Lambda_0+\mathcal{C}'_1K(t)^2~,
\end{equation}
where $\mathcal{C}_1$ is an integration constant and $\mathcal{C}'_1=144\mathcal{C}_1$.

This means the contribution of the cosmological term differs from the one arising from the vacuum energy, $\Lambda_0$.

 Furthermore, one has to find the conditions for the vector field such that the homogeneity and isotropy of the Universe are not spoiled. Thus, equations (\ref{eqn:fieldandtrace1}) and (\ref{eqn:fieldandtrace2}) can be combined into:
\begin{equation}
4\left[R_{\mu\nu}+\bar{\bar{R}}_{(\mu\nu)}\right]\bar{\Theta}=g_{\mu\nu}\left[R+\bar{\bar{R}}\right]\bar{\Theta}~,
\end{equation}
where the Levi-Civita Riemann tensor can itself generate a space form manifold, hence its contractions give $R_{\mu\nu}=3\Lambda g_{\mu\nu}$ and $R=12\Lambda$, which cancel in both sides of the previous equation and leads to an equation for the vector field for the non-trivial case, $\bar{\Theta}\neq 0$:
\begin{equation}
g_{\mu\nu}\nabla_{\lambda}A^{\lambda}+g_{\mu\nu}\frac{1}{2}A_{\lambda}A^{\lambda}=2A_{\mu}A_{\nu}+4\nabla_{(\mu}A_{\nu)}~.
\end{equation}

A possible ansatz for the vector field is the following:
\begin{equation}
A_0=\xi(t)~,~A_i=\chi(t)\delta^a_{i}L_a~,
\end{equation}
which admits invariance under spatial $SO(3)$ transformations with generators $L_a$, and accounts for the homogeneity and isotropy of space. Both the time-time component and each diagonal space-space components lead to:
\begin{equation}
\dot{\xi}+\frac{1}{2}\xi^2+\frac{1}{2}\chi^2=0~,
\end{equation}
whilst the time-spatial components read:
\begin{equation}
\dot{\chi}+2\xi\chi=0~.
\end{equation}

Let us consider two non-trivial cases: $\xi=\xi_0=const.$ and $\xi=\chi$. The first leads to the solution:
\begin{equation}
\chi(t)=\chi_0 e^{-2\xi_0(t-t_0)}~,
\end{equation}
where $\chi_0$ and $t_0$ are integration constants. This means that the vector field gives an exponentially decreasing contribution. For the space form manifold this means that:
\begin{equation}
12K(t)=\left(12\Lambda+\frac{3}{2}\xi_0\right)-6\xi_0 \chi(t) -\frac{9}{2}\chi(t)^2 ~,
\end{equation}
which at late times is expected to be constant, $12K(t)\to 12K=12\Lambda+\frac{3}{2}\xi_0$.

The other case, $\xi=\chi$, leads to:
\begin{equation}
\chi(t)=\frac{\chi_0}{1+\chi_0 (t-t_0)}~,
\end{equation}
which for the space form constant means:
\begin{equation}
12K(t)=12\Lambda + 3 \partial^0 \chi(t)-\frac{3}{2}\times 2\chi^2(t)=12\Lambda~.
\end{equation}

In this case, the vector field does not contribute to the vacuum of the theory, and the curvature of the space form corresponds to the curvature of the Levi-Civita connection.

\section{Conclusions}\label{sec:conclusions}

Weyl gravity generalises GR by admitting non-metricity in the affine connection. Extensions of GR with general non-metricity features are known to significantly differ from its metric-compatible versions. Therefore, in this paper we have generalised Weyl Gravity taking into consideration a non-minimal coupling between matter and curvature. This yields second order metric field equations thanks to the constrain arising from the equations of motion of the vector field.

We find that space form manifolds are admitted in these theories, which means that non-metricity affects the vacuum. The curvature of the space form implies that the vector field has to obey to some symmetries. In particular, if $A_{\mu}$ satisfies a $SO(3)$ symmetry to preserve homogeneity and isotropy of the Universe, then two decaying solutions are found: one leads to a time evolving curvature which at late times is expected to be constant, and the other which leads to the well known constant curvature of GR. These solutions, likewise GR can be made compatible with observations though fine tuning.

It is worth mentioning that besides the discussed features, it is remarking that Weyl gravity has the interesting property of avoiding some known instabilities of fourth order theories given the constraint on the vector field of the connection, which effectively turns it into a second order theory.


\section*{Acknowledgments}

The work of C.G. is supported by Fundação para a Ciência e a Tecnologia (FCT)
under the grant SFRH/BD/102820/2014.



\end{document}